\documentclass[aps,twocolumn,amsmath,amssymb,floatfix]{revtex4}
\usepackage{epsfig,psfrag,subfigure}
\usepackage{graphicx,psfrag,subfigure}

\begin{document}
\title{Double point contact in Quantum Hall Line Junctions}
\author{Eun-Ah Kim}
\author{Eduardo Fradkin}

\affiliation{Department of Physics, University of Illinois at
Urbana-Champaign, 1110
W.\ Green St.\ , Urbana, IL  61801-3080, USA}
\date{\today}

\newcommand{\D}{\displaystyle}
\newcommand{\noi}{\noindent}
\newcommand{\F}{{\textrm F}}

\begin{abstract}

We show that multiple point contacts on a barrier separating two laterally coupled quantum Hall fluids induce Aharonov-Bohm (AB) oscillations in the tunneling conductance. These quantum coherence effects provide new evidence for the Luttinger liquid behavior of the edge states of quantum Hall fluids. 
For a two point
contact, we identify  coherent and incoherent regimes determined by the relative magnitude of their
separation and the temperature. We analyze both
regimes in the strong and weak tunneling amplitude limits as well as their temperature dependence. We 
find that the tunneling conductance should exhibit AB oscillations
in the coherent regime, both at strong and  weak
tunneling amplitude with the same period but with different functional form.
\end{abstract}
\maketitle



Two-dimensional electron fluids in large magnetic fields offer an ideal setting 
to study non-trivial quantum coherence effects in strongly interacting 
macroscopic systems. 
It is well known that the excitations supported by the 
edges of quantum Hall fluids provide an ideal window to study this new physics. 
Already there are a
number of very interesting experiments
~\cite{milliken96,chang96,grayson98,glattli,reznikov} 
which have uncovered the non-trivial
Luttinger liquid behavior of these edge states~\cite{wen90,kane92a}. 

Recently \textcite{kang00} used a new experimental setup in which two quantum Hall fluids are laterally coupled along an atomically precise barrier. In these experiments Kang and coworkers found that for filling factors $\nu \gtrsim 1$, and for some range of filling factors,  a pronounced zero bias conductance (ZBC) peak appears in the tunneling conductance of the device. (The same effect reappears for $\nu \gtrsim 2$.) Two alternative mechanisms have been proposed to explain these experiments: a) Landau level mixing induced by the barrier potential ~\cite{kang00,mitra01,kollar02}, and b) tunneling at isolated quantum point contacts~\cite{kim03}.

In Ref.~\cite{kim03}, we showed that the salient features of the experiment of \textcite{kang00} can be successfully explained by modeling the system
as a pair of (coupled) chiral Luttinger liquids (the edge states on each sides of the barrier) in the presence of a single point contact (PC). 
In particular we showed that 
inter-edge Coulomb
interaction yields an effective reduced Luttinger 
parameter $K<1$ and that for $\nu \gtrsim 1$, the system crosses over to the strong tunneling amplitude
regime, leading to the appearance of zero-bias peak in
the tunneling conductance with a peak value at $T=0$ of
$G_t=Ke^2/h$. 
This crossover is controlled by the 
energy scales of this system: the bias voltage $V$, the crossover scale $T_K$, 
and the temperature $T$ (and the possibility by a small spin
polarization for $\nu\sim2$). We also predicted an increase in
the height of the zero-bias conductance (ZBC) peak for $T \lesssim T_K$.

However, if the barrier contains more than one tunneling center (as it surely does) 
a number of interesting quantum coherence effects must take place, and it is of interest
to investigate quantum coherence effects of multiple impurities and
their competition with thermal fluctuations. An example of effects of this sort has been considered some time ago by \textcite{chamon-freed-kivelson-sondhi-wen97}, who proposed a quasi-particle interference 
experiment based on a two-tunneling center device in the fractional
quantum Hall regime, as a way to measure directly the fractional statistics of Laughlin 
quasi-particles. However, in the FQH regime, only the case of weak
tunneling centers needs to be considered since in this regime tunneling at a point contact is an irrelevant perturbation. Instead, for $\nu \gtrsim 1$, the the system is 
in the strong tunneling limit for $T \lesssim T_K$, which is not accessible by perturbation theory. In this regime, tunneling processes become dominant, and an instanton expansion is required to describe the physics. This problem is closely related to that of scattering centers in quantum wires, first discussed by \textcite{kane92a,kane92b}. Our analysis closely follows their approach.

In this paper, we analyze the quantum Hall Line junction with two PC's both in the
strong and weak tunneling amplitude limits.
We show that the two-PC system may be in a coherent or in an incoherent regime
depending on the distance $a$ separating the tunneling centers. In the coherent regime 
the system exhibits Aharonov-Bohm (AB) oscillations in the form of a series of resonant tunneling processes in the strong tunneling 
limit. Instead, the  AB effect in the weak
tunneling limit has a simple sinusoidal form. In contrast, in the incoherent regime,  the strong and
weak tunneling  limits are related by duality. Naturally, a realistic barrier must contain more than two PC's, which will result in a more complex structure of AB oscillations than what we find for just two PC's. Nevertheless it is also natural to expect that as the temperature is lowered this pattern will reveal itself step by step with the strongest PC's giving rise to the most prominent features of the interference pattern. The observation of this AB interference pattern will provide a way to sort out whether the ZBP observed by \textcite{kang00} is due to Landau level mixing or to PC tunneling, since the former mechanism predicts a smooth non-periodic dependence of the tunneling conductance on the magnetic field.
 
\begin{figure}[h!]
\begin{center}
\psfrag{d}{$d$}\psfrag{a}{$a$}
\psfrag{B}{$\vec{B}$}
\includegraphics[width=.2\textwidth]{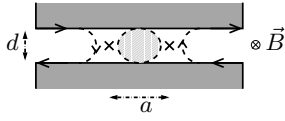}
\end{center}
\caption{A line junction with two tunneling centers. The two shaded
  regions  represent
  two identical 2DEG, separated by an insulating barrier mimicking  the
  junctions used in Ref. \cite{kang00}.}
\label{fig:setup}
\end{figure}

We begin by describing our model
(see Fig.\ref{fig:setup}) 
which has two PC's, one
located at $x=-a/2$ and the other located at $x=a/2$. In the notation of
Ref.~\cite{kim03}, the local tunneling operator is
\begin{equation}
{\mathcal H}_t=t_1\psi_+^\dagger \psi_-\delta(x+a/2)+
t_2\psi_+^\dagger \psi_-\delta(x-a/2) + h.c., 
\end{equation}
where $t_1$ and $t_2$ are tunneling amplitudes, hereafter referred to as the ``coupling constants". The right and left
moving chiral Fermi fields $\psi_\pm^\dagger$ can be bosonized~\cite{bosonization} in terms of the right and left moving chiral bosons $\phi_\pm$, as
$\psi_\pm^\dagger(x)\propto \frac{1}{\sqrt{2\pi}}
e^{\D\pm i \phi_\pm (x) \pm ik_\F x}$. Similarly, the normal-ordered density operators are given by 
$J_\pm = -\frac{1}{2\pi}\partial_x \phi_\pm$.
Notice that the tunneling operators at  $x=\pm \; a/2$ have a relative phase of $2k_\F a$, which
cannot be ``gauged away" by shifting the
bosonic field $\phi=\phi_+ +\phi_-$ by a constant.
However, the Fermi momentum is directly
connected to the position of the edge or the effective width $d$ of the
barrier and magnetic field through $k_\F=d/2{\ell}^2$ where ${\ell}=
\sqrt{\hslash c/eB}$ is magnetic length. Hence, the relative phase $2k_\F a$
is actually the Aharanov-Bohm (AB) phase 
$2k_\F a = 2\pi \Phi/\phi_0$,
 where $\Phi$ is the flux in the area enclosed by
tunneling edge currents and $\phi_0$ is the flux quantum. We now follow Ref. \cite{kim03} and
introduce the rescaled boson $\varphi=\phi/\sqrt{K}$, where $K=\sqrt{(1-g_c)/(1+g_c)}$ is the Luttinger parameter for the inter-edge 
interaction. In imaginary time, the total Lagrangian density is 
\begin{eqnarray}
{\mathcal L}&\!\!=\!\!&\frac{1}{8\pi}\left[\frac{1}{v}(\partial_\tau\varphi)^2
\!+\!v(\partial_x\varphi)^2\right]
\nonumber\\
&&+\sum_{\sigma=\pm} \Gamma_\sigma \D\cos\left[\smash{\sqrt{K}}\varphi\!-\sigma \; \pi\frac{\Phi}{\phi_0}\right]\delta(x\!+\sigma \;\frac{a}{2})  
\label{eq:fullL}
\end{eqnarray}
Since the tunneling perturbation only acts at the points $x=\pm a/2$, and the free
Luttinger liquid action is quadratic, we can integrate out
$\varphi(x)$ and write an effective action for the tunneling center
degrees of freedom $\varphi(\pm a/2,\tau)$. Let us 
introduce new variables $X_1\equiv(\varphi(-a/2,\tau)+\varphi(a/2,\tau))/\sqrt{2}$,
$X_2\equiv(\varphi(-a/2,\tau)-\varphi(a/2,\tau))/\sqrt{2}$, 
and to consider for now the case when
two impurities have the same strength 
$\Gamma_+ = \Gamma_- = \Gamma$. The effective action for $X_1(\tau)$ and $X_2(\tau)$ is
\begin{eqnarray}
S_{\rm eff}=&&\!\!\!\nonumber \\
\frac{1}{\beta}\!\sum_{n\neq0}&&\!\!\!\left[
\frac{|\omega_n|}{4\pi(1\!+\!e^{-|\omega_n|a})}|X_n^{(1)}|^2
+\frac{|\omega_n|}{4\pi(1\!-\!e^{-|\omega_n|a})}|X_n^{(2)}|^2
\right]\nonumber \\
+2\Gamma && \!\!\!\!\! \int_0^{1/T} \!\! \!\! d\tau  \cos\left(\!\sqrt{\frac{K}{2}}X_1\!\right)
\cos\left(\!\sqrt{\frac{K}{2}}X_2\!-\!\pi\frac{\Phi}{\phi_0}\!\right)
\label{eq:Seff-full}
\end{eqnarray}
where $\omega_n=2\pi n T$ are the Matsubara frequencies, 
$X_n^{(i)}$ are the Fourier components of
$X_i(\tau)$, and $T$ is the temperature.
Notice that $\frac{\sqrt{K}}{\pi2\sqrt{2}}X_1$ measures the {\it the
charge transferred along the barrier} and 
$\frac{\sqrt{K}}{\pi\sqrt{2}}X_2$ measures the {\it the
  charge transferred to the island} (the cross-hatched region in Fig.\ref{fig:setup}).

The term $e^{-\omega a}$ in the denominator of Eq. (\ref{eq:Seff-full})
accounts for the coherence between two tunneling centers. As a result, 
at sufficiently high temperatures $T a \gg 1$, $\exp(-\omega_n T) \ll 1$ and in this regime
quantum coherence effects are washed away. Thus, in physical units, we can identify two extreme regimes  in which  the
effective action simplifies: {\it the coherent regime} with $\hslash v/a\gg 
k_B T$ in which the PC's are strongly coupled, and {\it the incoherent regime} with $\hslash v/a\ll k_B T$, in which the PC's act independently. Thus, in the coherent regime, the effective
action of Eq. (\ref{eq:Seff-full}) 
becomes
\begin{eqnarray}
S_{\textrm{eff}}
= &&\!\!\!
\frac{1}{\beta}\sum_{n\neq0} \left[
\frac{|\omega_n|}{8\pi}|X_n^{(1)}|^2
+\frac{|\omega_n|}{8\pi}|X_n^{(2)}|^2
\right] 
\nonumber \\
&&+\int_0^{1/T} \!\!\!\!\! d\tau \; V_I(X_1,X_2)+ \ldots
\label{eq:Seff-island}
\end{eqnarray}
where $V_I$ is the effective potential 
\begin{eqnarray}
V_I(X_1,X_2)=&&\!\!\!\!\!\!\\
\frac{1}{4\pi a}X_2(\tau)^2 +&&\!\!\!\!\!2\Gamma\cos\left(\sqrt{\frac{K}{2}}X_1\right)
\cos\left(\sqrt{\frac{K}{2}}X_2-\pi\frac{\Phi}{\phi_0}\right)
\nonumber 
\label{eq:eff-pot}
\end{eqnarray}  
In contrast, in the incoherent regime   
$S_{\textrm eff}$ reduces to 
\begin{eqnarray}
S_{\textrm{eff}} &&\!\!\!=
\frac{1}{\beta}\sum_{n\neq0}\!\left[
\frac{|\omega_n|}{4\pi}|X_n^{(1)}|^2
+\frac{|\omega_n|}{4\pi}|X_n^{(2)}|^2
\right]
\nonumber \\
+2\Gamma&&\!\!\!\! \int_0^{1/T} \!\!\! d\tau\cos\left(\!\sqrt{\frac{K}{2}}X_1 \right)
\cos\left(\!\sqrt{\frac{K}{2}}X_2-k_\F a\!\right)
\label{eq:Seff-separate}
\end{eqnarray}
By comparing Eq. (\ref{eq:Seff-island}) and
Eq. (\ref{eq:Seff-separate}) we see that in the incoherent regime the PC's are effectively decoupled while in the coherent regime
they are strongly coupled, and 
$X_2$ becomes massive, with a mass of order $1/a$. Also note that the strength of
non-local interaction in the incoherent regime is exactly twice that of the
coherent regime. 

{\textbf {1) The coherent regime.}} 
The potential $V_I(X_1,X_2)$ is periodic in $X_1$ with period
$2\pi\sqrt{2/K}$. The mass term breaks this lattice translation
symmetry in $X_2$ direction. Thus, in general there exists a single value $X_2 =
X_2^0$ which minimizes the potential along the $X_2$ axis, unless the
resonance condition $\Phi/\phi_0=(\text{half-integer})$ is satisfied.
However, when the flux satisfies $\Phi=(n+\frac{1}{2})\phi_0$, 
the potential $V_I(X_1,X_2)$ acquires the additional symmetry
$V_I(X_1,X_2) = V(X_1+\pi\sqrt{2/K}, -X_2)$. Thus, in this case
 there are two values of $X_2$ which minimize the
potential. This effect is analogous to the resonance phenomena first pointed out by \textcite{kane92b} except for that in that case the
resonance was tuned by a gate voltage.
The resonance we found here is
the result of the Aharanov-Bohm effect which enables the transfer of
{\it half} an electron when the flux penetrating the island is exactly
half-integer flux quantum.  
We will use the instanton expansion~\cite{anderson-yuval-hamann70b,
leggett-chakravarty-dorsey-fisher-garg-zwerger,schmid83} to examine the coherent regime in the strong tunneling limit $\Gamma\gg1/(aK)$, and perturbation theory in the weak tunneling limit $\Gamma\ll1/(aK)$. 
\begin{figure}[ht!]
\centering
\mbox{
\psfrag{G}{\scriptsize $G_t$}\psfrag{B}{\scriptsize $B$}\psfrag{delB}{\scriptsize $\Delta
  B$}\psfrag{K}{\scriptsize $Ke^2/h$}
\psfrag{Gpert}{\scriptsize $G_t^{\textrm{pert}}$}\psfrag{delGres}{\scriptsize $\Delta
  G_t^{\textrm{res}}$}\psfrag{delGoff}{\scriptsize $\Delta G_t^{\textrm{off}}$}
\subfigure[weak tunneling amplitude limit]{\includegraphics[width=.2\textwidth]{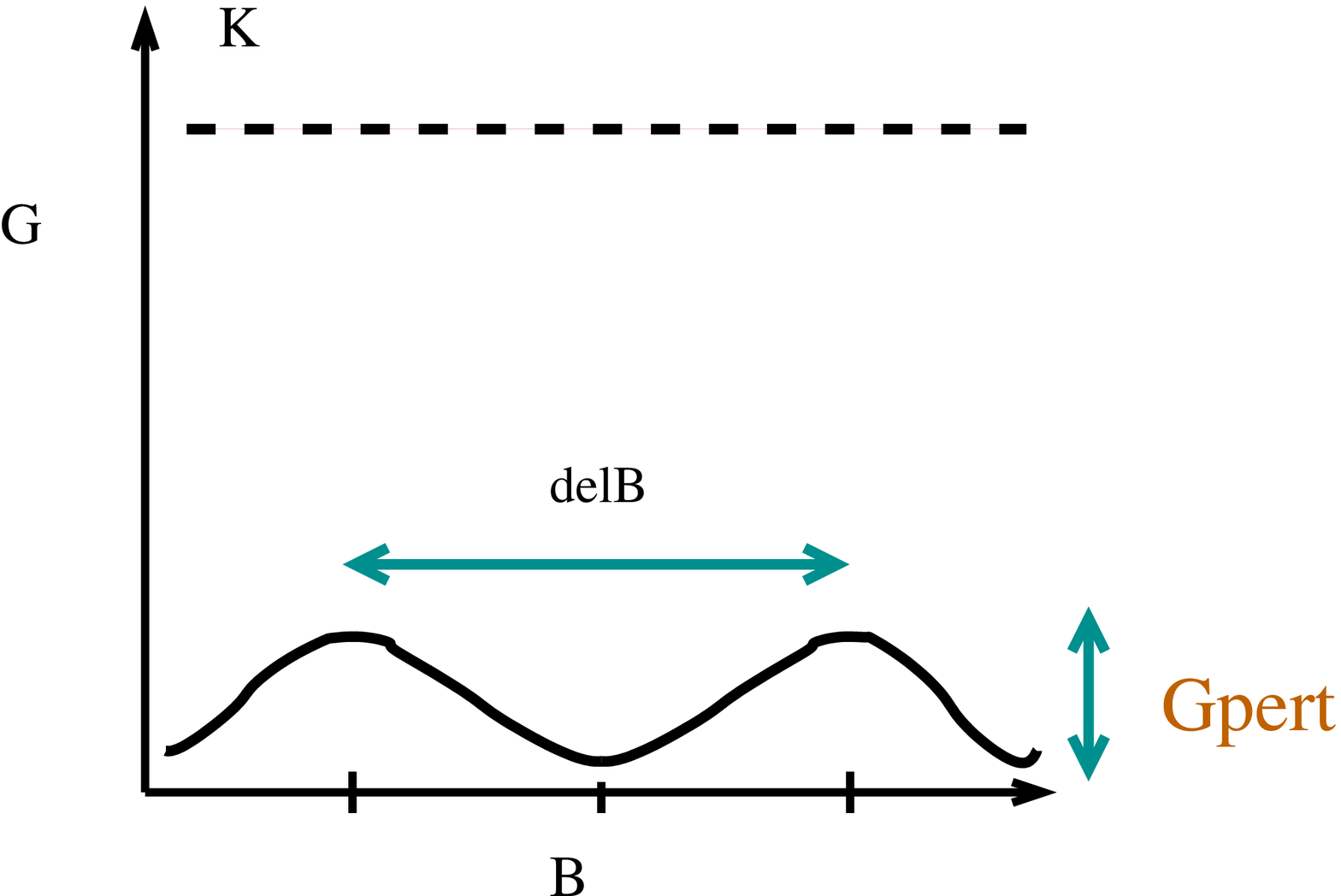}}\qquad
\subfigure[strong tunneling amplitude limit]{\includegraphics[width=.2\textwidth]{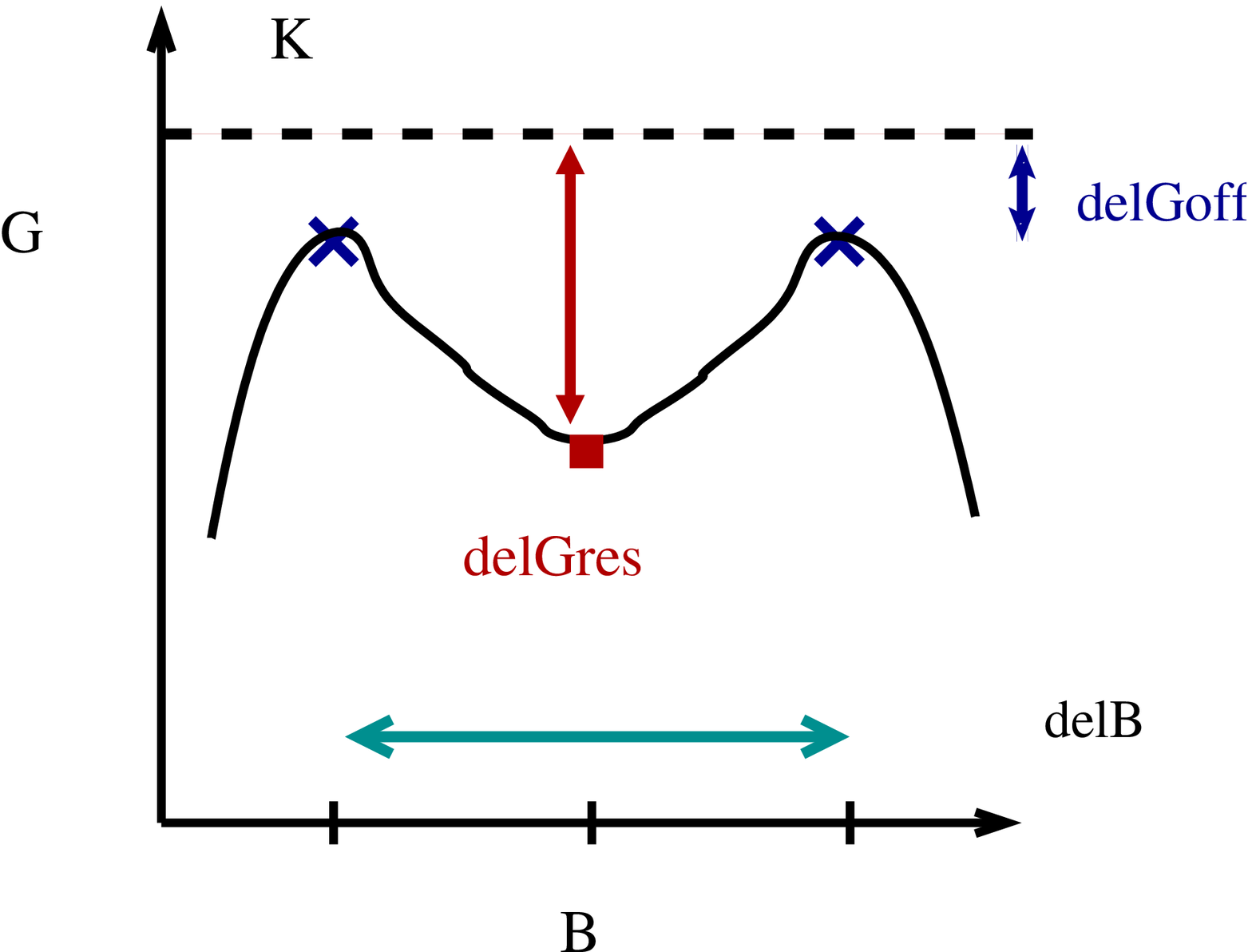}}}
\caption{(a) AB oscillations in the weak tunneling limit of
  the coherent regime; (b) AB effect in the strong tunneling limit
  of the coherent regime for $K<1/4$.}
\label{fig:ZBCP}
\end{figure}
{\it The weak tunneling limit}, $\Gamma\ll1/(aK)$: Here the effective potential  is
dominated by the mass term which is minimized for
$X_2=0$. Hence in this case $X_2$ can be integrated out
resulting simply in a  finite flux-dependent renormalization of the tunneling amplitude:
$V_I\rightarrow 2\Gamma\cos(\pi\frac{\Phi}{\phi_0})
\cos\left(\sqrt{\frac{K}{2}}X_1\right)$. Consequently, a lowest order
perturbative calculation using the Keldysh
formalism~\cite{kadanoff-baym,rammer86} results an expression for 
$G_t(V=0,T)$ in the coherent weak tunneling limit: 
\begin{eqnarray}
G_t(0,T)&=&K\frac{e^2}{h}
\frac{K}{2}\pi^{2K}
\frac{\Gamma(1/2)\Gamma(K)}{\Gamma(1/2+K)}\left(\frac{T}{T_K^{CW}}\right)^{(2K-2)}\nonumber\\
&&\times\frac{1}{2}(1+\cos(2\pi\Phi/\phi_0))+\cdots,
\label{eq:Gt-co-weak}
\end{eqnarray}
where $T_K^{CW}=\Lambda\left(\frac{2\Gamma}{\Lambda}\right)^{1/(1-K)}$ is a crossover scale and $\Lambda$ is a high-energy cutoff.
Hence, in the
weak tunneling limit of the coherent regime, there is an Aharonov-Bohm interference  effect with the 
usual oscillatory form (albeit with a reduced amplitude), as well as an offset . 

{\it The strong tunneling limit }: 
For $\Gamma\gg 1/(aK)$, the system can be
either off-resonance or on-resonance. When the system is
{\it off-resonance}, $V_I$ has a single minimum at
$X_2=X_2^0\approx\sqrt{2/K}(k_\F a-n\pi)$, where $n$ is an integer. In this case, there is an energy
gap of order $1/aK$ to the states with other values of $X_2$, {\it i.e.\/} the
island-charge fluctuation is effectively suppressed (Coulomb blockade). Thus,
$X_2$ is frozen to a non-zero value and the problem reduces to a
single point contact system. In this regime the saturation value of the tunneling
conductance is necessarily equal to $Ke^2/h$. We can
calculate the leading corrections to this result using the instanton technique.
In this case, the instanton is an electron tunneling process across the island between two
disconnected pieces of the barrier.
We will denote the instanton fugacity by $\zeta$, and compute
 the lowest order correction to the tunneling conductance
$\Delta G_t^{\textrm{off}}\equiv G_t-Ke^2/h$ due to these tunnelling processes. 
To the lowest order in $\zeta$ we find
\begin{equation}
\Delta G_t^{\textrm{off}}=-\frac{\pi^{\frac{2}{K}}}{4}
\frac{\Gamma(1/2)\Gamma(1/K)}{\Gamma(1/2+1/4K)}
\left(\frac{T}{T_K^{\textrm{CSO}}} \right)^{\frac{2}{K}-2}+\cdots,
\label{eq:delG-off}
\end{equation} 
where  $T_K^{\textrm{CSO}}=\Lambda\left(\frac{\Lambda}{\zeta}\right)^{K/(1-K)}$ is a crossover scale determined by the instanton fugacity
$\zeta$ and $\Lambda$. 

However, when the magnetic field is tuned to a
{\it resonance}, the projection of $V_I$
has two degenerate minima at $X_2=\pm
\frac{\pi}{2}\sqrt{\frac{2g}{K}}$, where $g$ represents the renormalization of the Luttinger parameter (or ``compactification radius") 
of $X_2$; at the
strong tunneling fixed point  $g_0=1$. Hence the island
effectively behaves like a two-level system and
instantons connecting degenerate minima $(X_1 = m\pi\sqrt{2/K},
X_2=X_2^0)$ and $(X_1 = (m\pm1)\pi\sqrt{2/K}, X_2=-X_2^0)$  with integer $m$ 
correspond to electrons hopping on and off the island with the
hopping amplitude $\zeta'$. 
In terms of these instantons, with
fugacity $\zeta'$,  the partition
function is \cite{kane92b,anderson-yuval-hamann70b}  
\begin{eqnarray}
{\mathcal Z} &=& \sum_n \!\!\sum_{q_i^{(\!1\!)}=\pm1}'\!\zeta'^{2n}\!
\int_0^\beta d\tau_{2n}\cdots\int_0^{\tau_2}
d\tau_1\;  e^{-\sum_{i<j}V_{ij}}\  ,\nonumber\\
V_{ij}=&&\!\!\!\!\!\!\!\!\!\!-\frac{1}{2K} 
(q_i^{(\!1\!)}q_j^{(\!1\!)}+gq_i^{(\!2\!)}q_j^{(\!2\!)})
\ln\left|\Lambda(\tau_i-\tau_j)\right|
\label{eq:instanton-expansion}
\end{eqnarray}

As long as the
$\zeta'/\Lambda\ll1$, we can use this partition function to
 calculate semi-classically the lowest order correction to $\Delta
G_t^{\textrm{res}}\equiv G_t^{\textrm{res}}-Ke^2/h$:
\begin{equation}
 \Delta G_t^{\textrm{res}}=-\frac{1}{4}\pi^{\frac{(1+g)}{2K}}
\frac{\Gamma((1+g)/4K)}{\Gamma(1/2+(1+g/4K))}
\left(\frac{T}{T_K^{\textrm{CSR}}}\right)^{\frac{(1+g)}{2K}-2}
\label{eq:delG-res}
\end{equation}
with $T_K^{\textrm{CSR}}=\Lambda(\frac{\Lambda}{\zeta'})^{\frac{4K}{1+g-4K}}$.
Comparing Eq. (\ref{eq:delG-off}) and Eq. (\ref{eq:delG-res}), we observe that the
tunneling conductance {\it on resonance} is  further suppressed than {\it off-resonance}. More explicitly, we find that
the {\it ratio} of the off and on resonance corrections obey the scaling law
\begin{equation}
\frac{\Delta G_t^{\textrm{res}}}{\Delta G_t^{\textrm{off}}}\propto
\left(\frac{1}{T}\right)^{(3-g)/2K}.
\label{eq:Gt-ratio}
\end{equation}

Thus, the  Aharonov-Bohm effect leads to an oscillatory behavior of the tunneling conductance both at strong and at weak  $\Gamma$. In the weak tunneling limit it leads to
a small amplitude sinusoidal oscillation of $G_t$. At strong tunneling, although the tunneling conductance is closer to its maximum value $Ke^2/h$, the deviations are more pronounced and are governed by a series of 
resonances. In particular, although quantum coherence involves different mechanisms in the weak and strong tunneling limits, the
periodicity is the same in both regimes. A simple estimate of the period is
$\Delta B\approx 0.2$ Tesla for two PC's separated by a distance
$a\approx100 \; {\ell}$ (see Fig.~\ref{fig:ZBCP}). 
 
\begin{figure}[ht!]
\centering
\psfrag{z}{\scriptsize $\zeta$}\psfrag{g}{\scriptsize $g$}\psfrag{4k-1}{{\scriptsize $4K\!-\!1$}}\psfrag{g0}{\scriptsize $g_0$}
\psfrag{0}{\scriptsize $0$}
\mbox{
\subfigure[]{\includegraphics[height=.15\textwidth]{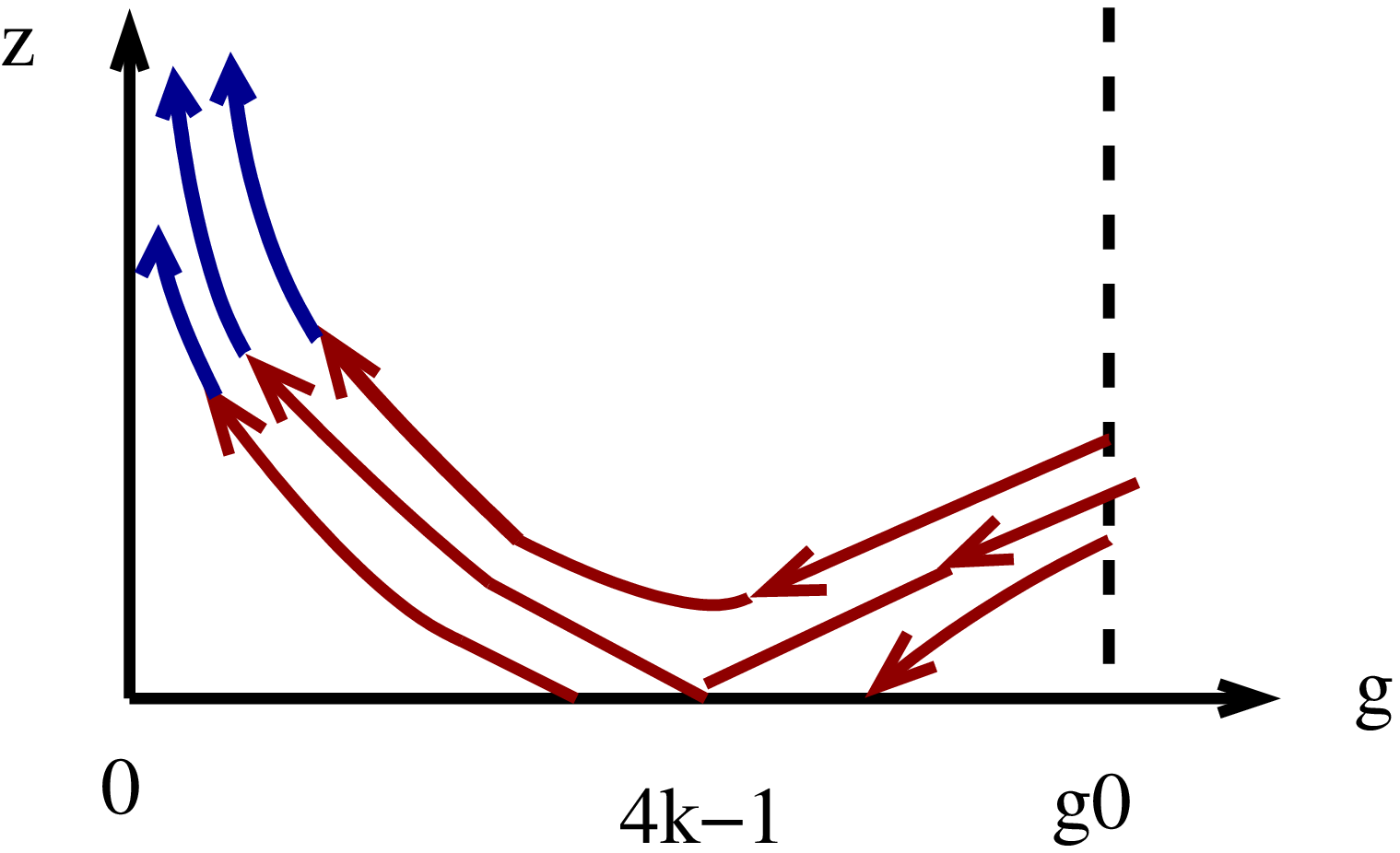}}\quad
\psfrag{z}{$\zeta$}\psfrag{g}{\scriptsize $g$}\psfrag{4k-1}{{\scriptsize $ $}}\psfrag{g0}{\scriptsize $g_0$}
\psfrag{0}{\scriptsize $0$}
\subfigure[]{\includegraphics[height=.15\textwidth]{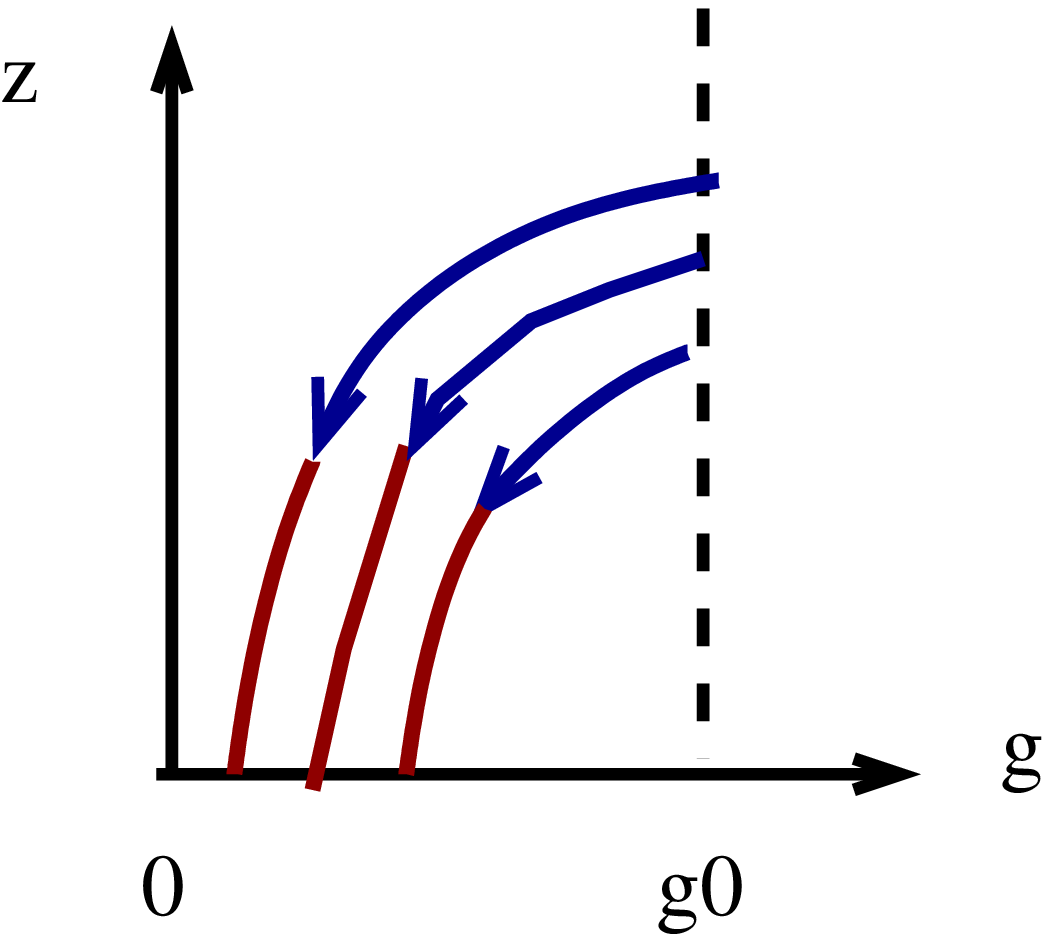}}
}
\caption{RG flow for (a) $1/4<K<1/2$ and (b) $K<1/4$}
\label{fig:phase}
\end{figure}
{\it Phase transition or crossover?} Having understood the strong and the weak tunneling
limits in the coherent regime, we now inquire if these regimes are separated by a phase transition or by a smooth
cross-over. The RG flow for this problem, originally derived in Ref. 
\cite{kane92b}, is:
\begin{subequations}
\label{eq:RG}
\begin{eqnarray}
\frac{\delta\zeta'}{\delta l} &=&\frac{1}{4K}((4K-1)-g)\zeta'\\ 
\frac{\delta K}{\delta l} &=&-\frac{8}{\Lambda^2}\zeta'^2g.
\end{eqnarray}
\end{subequations}

This resulting flow, shown qualitatively in Fig. \ref{fig:phase}, has  a  fixed point at $(g,\zeta')=(4K-1,0)$ which depends
on $K$, and on the initial value of
$g$, which is $g_0=1$. The asymptotic behavior of the system depends on both the
value of $K$ and the initial value of $\zeta'$. Thus, for $1/4\leq K<1/2$, there is a quantum
phase transition at $T=0$ between a phase in which the tunneling conductance $G_t$ saturates to $Ke^2/h$ as $T \to 0$, and a phase in which $G_t$ vanishes as $T \to 0$.
Instead, for $K<1/4$, there is a crossover as
the system flows to a line of strong tunneling fixed
points $g\!\to\!g*$, $\zeta'\!\to0$, each of which yielding a different
scaling law for Eq. (\ref{eq:Gt-ratio}).
For $T>0$,
in all cases, there will be a crossover between strong and weak
tunneling fixed points as finite temperature will eventually stop the flow. The data of Ref. \cite{kang00} suggests that Coulomb interactions reduces the Luttinger parameter
to a small value $K\sim 0.2$.    
 
{\textbf {2) The incoherent regime}}: In this regime,  $\exp(-\beta a) \to 0$, thermal fluctuations overwhelm coherence
effects and the two PC's behave as if they were decoupled from each other. 
To the lowest order in $\Gamma/\Lambda$, the zero bias tunneling conductance in the
weak tunneling limit $T\gg \Lambda$ calculated perturbatively is 
\begin{equation}
G_t=K\frac{e^2}{h}-\pi^{2/K}
\frac{\Gamma(1/2)\Gamma(1/K)}{\Gamma(1/K+1/2)}
\left(\frac{T}{T_K^{\textrm{IS}}}\right)^{(2/K-2)}
\!\!\!\!\!\! +\ldots
\label{eq:incoherent-weak-Gt}
\end{equation}
 with $T_K^{\textrm{IS}}=\Lambda\left(\frac{\Lambda}{\zeta}\right)^{\frac{K}{1-K}}$. Similarly, a semi-classical calculation in the strong tunneling limit
$T\ll \Lambda$ leads to
\begin{equation}
G_t=\frac{e^2}{h}\frac{K\pi^{(2K)}}{2}\left(\frac{\Gamma}{\Lambda}\right)^2
\frac{\Gamma(1/2)\Gamma(K)}{\Gamma(K+1/2)}\left(\frac{T}{T_K^{\textrm{IW}}}\right)^{(2K-2)}
\!\!\!\!\!\!\! +\ldots
\label{eq:incoherent-strong-Gt}
\end{equation}
with  $T_K^{\textrm{IW}}=\Lambda\left(\frac{\Gamma}{\Lambda}\right)^{\frac{1}{1-K}}$.
Eq.(\ref{eq:incoherent-weak-Gt}) and
Eq.(\ref{eq:incoherent-strong-Gt}) show that, in the incoherent
regime, weak and strong
tunneling limits are exactly dual to each other. They also have the same scaling
behavior as the single PC case studied in Ref.~\cite{kim03}, which
explains why the single PC picture works.  

We close with a few comments on multi-impurity effects. Here we have discussed in detail coherence effects in a two PC system. Naturally, a realistic barrier (even an ``atomically precise" one) will have a number of such defects. It is clear that a multi-point contact extension of our analysis will lead to a complex interference pattern due to the existence of many competing pathways. Also, one expects a broad distribution of defects, both in tunneling amplitudes and in relative separation. Thus, at a given temperature, the strongest effects will be due to the closest defects with the largest tunneling amplitudes. Thus, as $T$ is lowered, coherent Aharonov-Bohm oscillations will become increasingly more complex. Conversely, as $T$ is raised these effects are washed-out and the system will eventually reach the single impurity limit. 

After this work was completed we became aware of the experiments of \textcite{kataoka02} on a anti-dot in a quantum Hall system. Although the main focus of that work is the behavior of the system near $\nu=2$ (which we will discuss elsewhere) we wish to note that the device of Ref. \cite{kataoka02} is in the strong tunneling coherent regime discussed above. Further, the temperature dependence of the oscillations observed in Ref. \cite{kataoka02}  are remarkably similar to what we find for Aharonov-Bohm oscillations here near $\nu=1$. Also, \textcite{sim03} have recently developed a model for the anti-dot device.  

We thank Prof.\ W.\ Kang for several useful and stimulating discussions.  
This work was supported in part by the National
Science Foundation through the grant DMR01-32990.

\bibliography{journal,tunneling}
\end{document}